\documentclass[12pt]{aastex}

\slugcomment{Draft: gj876\_rev11.tex}

\begin{document}
\title{The Lick-Carnegie Exoplanet Survey: A Uranus-mass Fourth Planet for GJ~876 in an Extrasolar Laplace Configuration\altaffilmark{1}}

\bigskip
\author{
Eugenio J. Rivera\altaffilmark{2},
Gregory Laughlin\altaffilmark{2},
R. Paul Butler\altaffilmark{3},
Steven S. Vogt\altaffilmark{2},
Nader Haghighipour\altaffilmark{4},
Stefano Meschiari\altaffilmark{2}
}

\email{rivera@ucolick.org}

\altaffiltext{1}{Based on observations obtained at the W.M. Keck Observatory,
which is operated jointly by the University of California and the California
Institute of Technology.}

\altaffiltext{2}{UCO/Lick Observatory, University of California at Santa Cruz,
Santa Cruz, CA 95064, USA}

\altaffiltext{3}{Department of Terrestrial Magnetism, Carnegie Institution
of Washington, 5241 Broad Branch Road NW, Washington DC, 20015-1305, USA}

\altaffiltext{4}{Institute for Astronomy \& NASA Astrobiology Institute,
University of Hawaii-Monoa, 2680 Woodlawn Drive, Honolulu, HI 96822, USA}

\begin{abstract}

Continued radial velocity monitoring of the nearby M4V red dwarf star GJ~876
with Keck/HIRES has revealed the presence of a Uranus-mass fourth planetary
companion in the system.  The new planet has a mean period of $P_e=126.6$ days
(over the 12.6-year baseline of the radial velocity observations), and a
minimum mass of $m_e\sin{i_e}=12.9\pm 1.7\,M_{\oplus}$. The detection of the new
planet has been enabled by significant improvements to our radial velocity data
set for GJ~876. The data have been augmented by 36 new high-precision
measurements taken over the past five years. In addition, the precision of all
of the Doppler measurements have been significantly improved by the
incorporation of a high signal-to-noise template spectrum for GJ~876 into the
analysis pipeline. Implementation of the new template spectrum improves the
internal RMS errors for the velocity measurements taken during 1998-2005 from
4.1 m\,s$^{-1}$ to 2.5 m\,s$^{-1}$.  Self-consistent, N-body fits to the
radial velocity data set show that the four-planet system has an invariable
plane with an inclination relative to the plane of the sky of $i=59.5^{\circ}$.
The fit is not significantly improved by the introduction of a mutual
inclination between the planets ``b'' and ``c,'' but the new data do confirm a
non-zero eccentricity, $e_d=0.207\pm0.055$ for the innermost planet, ``d.'' In
our best-fit coplanar model, the mass of the new component is
$m_e=14.6\pm1.7\,M_{\oplus}$.  Our best-fitting model places the new planet in a
3-body resonance with the previously known giant planets (which have mean
periods of $P_c=30.4$ and $P_b=61.1$ days).  The critical argument,
$\varphi_{\rm Laplace}=\lambda_c-3\lambda_b+2\lambda_e$, for the Laplace
resonance librates with an amplitude of
$\Delta\varphi_{\rm Laplace}=40\pm13^{\circ}$ about
$\varphi_{\rm Laplace}=0^{\circ}$.  Numerical integration indicates that the
four-planet system is stable for at least a billion years (at least for the
coplanar cases).  This resonant configuration of three giant planets orbiting an
M-dwarf primary differs from the well-known Laplace configuration of the three
inner Galilean satellites of Jupiter, which are executing very small librations
about $\varphi_{\rm Laplace}=180^{\circ}$, and which never experience triple
conjunctions.  The GJ~876 system, by contrast, comes close to a triple
conjunction between the outer three planets once per every orbit of the outer
planet, ``e.''

\end{abstract}


\section{Introduction}
\label{intro}

The planetary system orbiting the nearby M4V star GJ~876 (HIP 113020) has proven
to be perhaps the most remarkable entry in the emerging galactic planetary
census.  This otherwise unassuming red dwarf has produced the first example of a
giant planet orbiting a low-mass star, the first instance of a mean-motion
resonance among planets, the first clear-cut astrometric detection of an 
extrasolar planet, and one of the first examples of a planet in the hitherto
unknown mass regime between Earth and Uranus.

The radial velocity (RV) variations of GJ~876 have been monitored by us, the
Lick-Carnegie Exoplanet Survey team (LCES), at the Keck I telescope using the
High Resolution Echelle Spectrograph (HIRES) for 12.6 years.  \citet{BS09} and
\citet{R05} (R05 henceforth) give detailed discussions of the discoveries and
some of the studies made for this system.  Here, we summarize some of that
history.

\citet{Marcy98} and \citet{detal98} announced the first companion, ``b.''  They
found that it has an orbital period ($P_b$) of $\sim 61$ days and a minimum mass
($m_b\sin{i_b}$) of $\sim 2.1\,M_{\rm Jup}$ and that it produced a reflex
barycentric velocity variation of its host star of amplitude
$K_b \sim 240$ m\,s$^{-1}$.  After 2.5 more years of additional Doppler
monitoring, \citet{Marcy01} announced the discovery of a second giant planet in
the system.  This second companion, ``c,'' has an orbital period, $P_c\sim30$
days, $m_c\sin{i_c} \sim 0.56\,M_{\rm Jup}$, and $K_c \sim 81$ m\,s$^{-1}$, and
upon discovery of the second planet, the parameters of the first planet were
tangibly revised.  \citet{Marcy01} modeled the two-planet system with
non-interacting Keplerian orbits, and noted both that there was room for
improvement in the quality of the fit and that the system's stability depended
on the initial positions of the planets.

To a degree that has not yet been observed for any other planetary system, the
mutual perturbations between the planets are dynamically significant over
observable time scales.  This fortuitous situation arises because the minimum
masses are relatively large compared to the star's mass, because the planetary
periods are near the 2:1 commensurability, and because their orbits are confined
to a small region around the star.

\citet{LC01} and \citet{RL01} independently developed self-consistent
``Newtonian'' fitting schemes which incorporate the mutual perturbations among
the planets in fitting the RV data. The inclusion of planet-planet interactions
resulted in a substantially improved fit to the RV data, and suggested that
the coplanar inclination of the b-c pair lies near $i=50^{\circ}$.  These papers
were followed by an article by \citet{N02} who described a similar dynamical
fitting method, but, in contrast, found a system inclination near
$i=90^{\circ}$.

Soon after the announcement of planet ``c,'' \citet{astrometry} used the Fine
Guidance Sensor on the Hubble Space Telescope to detect the astrometric wobble
induced by planet ``b,'' which constituted the first unambiguous astrometric
detection of an extrasolar planet. Their analysis suggested that the orbital
inclination of planet ``b'' is close to edge-on ($i_{b}=84^{\circ}\pm6^{\circ}$),
a result that agreed  with the model by \citet{N02}, but which was in conflict
with the results of \citet{LC01} and \citet{RL01}.

R05 analyzed an updated RV data set that included new Doppler velocities
obtained at Keck between 2001-2005.  By adopting self-consistent fits, they were
able to announce the discovery of a third companion, ``d,'' with period
$P_d\sim1.94$ days, $m_d\sin{i_d} \sim 5.9\,M_{\oplus}$, and
$K_d \sim 6.5$ m\,s$^{-1}$.  Using only the RV data, they were able to constrain
the inclination of the system (assuming all three planets are coplanar) relative
to the plane of the sky to $50\pm3^{\circ}$.  For this inclination, the true
mass of the third companion is $\sim 7.5\,M_{\oplus}$.

\citet{BS09} performed self-consistent fitting of both the Keck RV data from R05
and the astrometry from \citet{astrometry}.  They found the that the joint
data set supports a system inclination $i=48.9^{+1.8^{\circ}}_{-1.6^{\circ}}$, which
is in agreement with the $i=50\pm3^{\circ}$ value published by R05. Bean \&
Seifahrt's analysis (see, e.g. their Figure 1) indicates that the best-fit
$\chi_{\nu}^2$ is not significantly affected by the inclusion of the
astrometric data, and that the astrometric detection of ``b'' by
\citet{astrometry} is not in conflict with the significantly inclined system
configuration. Bean \& Seifahrt's work also resulted in the first determination
of a mutual inclination between the orbits of two planets in an extrasolar
planetary system.  By allowing the inclinations, $i_b$, and $i_c$, and the
nodes, $\Omega_b$, and $\Omega_c$ to float as free parameters in their
three-planet dynamical fit, they derived a mutual inclination,
$\cos{\Phi_{bc}}=\cos{i_b}\cos{i_c}+\sin{i_b}\sin{i_c}\cos{(\Omega_c-\Omega_b)}$,
of $\Phi_{bc}=5.0^{+3.9^{\circ}}_{-2.3^{\circ}}$.

\citet{HARPS} also performed self-consistent, mutually inclined fits using the
RVs from R05 plus 52 additional high precision RV measurements taken with the
HARPS spectrograph.  They confirmed the presence of companion ``d,'' and further
showed that it has a significant orbital eccentricity, $e_d=0.139\pm0.032$.
\citet{HARPS} determined inclinations $i_b=48.93^{\circ}\pm 0.97^{\circ}$ 
and $i_b=48.07^{\circ}\pm 2.06^{\circ}$. Their orbital fit yields a mutual
inclination $\Phi_{bc}=1.00^{\circ}$, consistent to within measurement
uncertainty with a coplanar system configuration.  We became aware of
\citet{HARPS} while preparing this article for publication. A. Correia (2009,
personal communication) graciously shared the HARPS radial velocity data on
which their paper is partly based.  We have carried out a preliminary analysis
that indicates that our new results are not in conflict with these HARPS data.
The HARPS RVs do not add much to constrain the system's dynamics because of
poor phase coverage at the period of the forth planet to be discussed below.
Throughout this article, however, our analysis and results are based on the Keck
RVs alone.

In addition to R05, several other studies, such as \citet{West99},
\citet{West01}, Laughlin et al. (2005), and \citet{Shankland06}, have examined
the possibility of transits occurring in this system.  The star has also been
included in surveys in which it was probed for nearby companions or a
circumstellar dust disk.  These include \citet{Leinert97}, \citet{Patience02},
\citet{Hinz02}, \citet{Trilling00}, \citet{Lestrade06}, \citet{Shankland08}, and
\citet{Bryden09}.  \citet{LJ02} were able to use Keck AO to place limits on
potential brown dwarfs in the system at separations of 1\,--\,10 AU.
\cite{SD09} used the Spitzer spacecraft to search for thermal emission from
planet ``d.''  The star
was also included in the Gemini Deep Planet Survey \citep{Lafreniere07}.  In
addition to some of the work above, a variety of studies have examined the
dynamics, and/or long-term stability of the system.  Examples are \citet{KN01},
\citet{Goz01} and \citet{Goz02}, \citet{Snellgrove01}, \citet{Ji02},
\citet{Jones01}, \citet{Zhou03}, and \citet{Nader03}.  \citet{BM03} present an
analytic model which does a very good job at reproducing the dynamical evolution
of the system.  \citet{Murray02}, \citet{LP02}, \citet{Chiang02}, \citet{TL03},
\citet{BFM03}, \citet{KPB04}, \citet{Lee04}, \citet{Kley05}, and \citet{LT09}
examined scenarios of how the system may have come to be in its current
dynamical configuration.  Even interior structure models of GJ~876\,d have been
introduced \citep{Valencia07}.

In this work, we present a detailed examination of our significantly augmented
and improved HIRES RV data set from the Keck observatory.  Our analysis
indicates the presence of a fourth, Uranus-mass planet in the GJ~876 system.
The new planet is in a Laplace resonance with the giant planets ``b'' and ``c,''
and the system marks the first example of a three-body resonance among
extrasolar planets. We expect that the newly resolved presence of the fourth
planet will provide additional constraints on the range of formation scenarios
that could have given rise to the system, and will thus have a broad impact on
current theories for planetary formation.
The plan for this paper is as follows:  In \S~\ref{obs}, we describe
the new velocities and the methods used to obtain them.  In \S~\ref{2plcf}, we
present our latest two- \& three-planet coplanar fits.  In \S~\ref{Res-3pl}, we
thoroughly analyze the residuals of our best three-planet coplanar fits.  In
\S~\ref{4plcf}, we discuss our four-planet fits.  In \S~\ref{Laplace} we explore
the possibility of constraining the libration amplitude of the critical angle
for the Laplace resonance.  In \S~\ref{morepls} we explore the stability of
additional planets in the system.  Finally, in \S~\ref{dis}, we finish with a
summary and a discussion of our findings.

\section{Radial Velocity Observations}
\label{obs}

The stellar characteristics of GJ~876 have been described previously in
\citet{Marcy98} and \citet{Laugh05}.  It has a Hipparcos distance of 4.69 pc
\citep{petal97}.  Its luminosity is 0.0124 $L_{\odot}$.  As in previous studies,
we adopt a stellar mass of $0.32\,M_{\odot}$ and a radius of $0.3\,R_{\odot}$
based on the mass-luminosity relationship of \citet{HM93}.  We do not
incorporate uncertainties in the star's mass ($0.32\,\pm\,0.03\,M_{\odot}$) into
the uncertainties in planetary masses and semi-major axes quoted herein.  The
age of the star exceeds 1 Gyr \citep{Marcy98}.

We searched for Doppler variability using repeated, high resolution spectra
with resolving power $R\approx70000$, obtained with Keck/HIRES \citep{Vogt94}.
The Keck spectra span the wavelength range from 3700\,--\,8000 \AA.  An iodine
absorption cell provides wavelength calibration and the instrumental profile
from 5000 to 6200 \AA \ \citep{MB92,Butler96}.  Typical signal-to-noise ratios
are 100 per pixel for GJ~876.  At Keck we routinely obtain Doppler precision of
better than 3\,--\,5 m\,s$^{-1}$ for V\,=\,10 M dwarfs.  Exposure times for
GJ~876 and other V\,=\,10 M dwarfs were typically 8 minutes. Over the past year,
these typical exposure times have been raised to 10 minutes.

The internal uncertainties in the velocities are judged from the velocity
agreement among the approximately 700 2-\AA\, chunks of the echelle spectrum,
each chunk yielding an independent Doppler shift.  The internal velocity
uncertainty of a given measurement is the uncertainty in the mean of the
$\sim 700$ velocities from one echelle spectrum.

We present results of N-body fits to the RV data taken at the W. M. Keck
telescope from 1997 June to 2010 January.  The 162 measured RVs are listed in
Table~\ref{velocities}.  The median of the internal uncertainties is
2.0 m\,s$^{-1}$.  Comparison of these velocities with those presented in
\citet{Laugh05} and in R05 shows significant changes (typically
3\,--\,10 m\,s$^{-1}$) in the velocities at several observing epochs.  R05
discuss the primary reasons for the improvements --- a CCD upgrade and an
improved data reduction pipeline.  Additionally, many of the velocities
presented here are the result of averaging multiple exposures over two-hour time
bins.  More importantly, the present RV data set is based on a new ``super''
template.  For the new template, ten exposures, most of which were $<900$ sec,
were obtained under very good observing conditions.  In contrast, the template
used to determine the RVs and uncertainties in R05 was based on only three
900-sec exposures taken under relatively poor observing conditions.  As a
result, the new template has brought the median internal uncertainty down to the
current 2.0 m\,s$^{-1}$ from the value of 4.1 m\,s$^{-1}$ in R05.  Additionally,
the new template has resulted in significant changes in the old published RVs.
Note that since the last observation listed in R05, 36 more high quality
observations have been obtained for this system since 2004 December.  Our LCES
survey assigns high observing priority to GJ~876, given the potential for
further characterizing the intricate dynamical configuration between the 30-
and 61-day planets and because of the possibility of detecting additional
planets in the system.

\begin{deluxetable}{rrr}
\tabletypesize{\scriptsize}
\tablecaption{Measured Velocities for GJ~876 (Keck)}
\tablewidth{0pt}
\tablehead{
BJD$~~$ & RV$~~$ & Unc.$~~$ \\
(-2450000)   &  (m\,s$^{-1}$) & (m\,s$^{-1}$)
}
\startdata
 602.09311 &   329.19 &  1.79\\ 
 603.10836 &   345.30 &  1.81\\ 
 604.11807 &   335.99 &  1.92\\ 
 605.11010 &   336.00 &  1.93\\ 
 606.11129 &   313.94 &  1.88\\ 
\enddata
\tablecomments{Table 1 is presented in its entirety in the electronic edition
of the Astrophysical Journal.  A portion is shown here for guidance regarding
its form and content.}
\label{velocities}
\end{deluxetable}

\section{Two- \& Three-Planet Coplanar Fits}
\label{2plcf}

We used similar procedures to those described in R05 to produce updated two-
and three-planet dynamical fits to the GJ~876 Doppler velocities.  The epoch
for all the fits in this work is JD 2450602.093.  As in R05, we initially held
the eccentricity of companion ``d'' ($e_d$) fixed. We find, however, that
fitting for $e_d$ results in a significant improvement in all the fits which
include companion ``d.'' We therefore allow $e_d$ to float in our final fits.
Uncertainties in the tables listing our best-fit jacobi parameters are based on
1000 bootstrap realizations of the RV data set.  We take the uncertainties to be
the standard deviations of the fitted parameters to the 1000 trials.  The
fitting algorithm is a union of Levenberg-Marquardt (LM) minimization (\S 15.5
of \citet{press}) and the Mercury integration package \citep{Cha99}.

Our dynamical fits can be integrated to determine the libration amplitudes of
the critical arguments of the 2:1 MMR between ``b'' and ``c,'' and to monitor
the linear secular coupling between the pericenter longitudes.  For each fit, we
integrated the system for 300,000 days ($\sim$821 years) with a timestep of
0.05 day.  For all simulations in this work, we used the Hybrid symplectic
algorithm in the Mercury integration package \citep{Cha99} modified as in
\citet{LR01} to include the partial first order post-Newtonian correction to
the central star's gravitational potential.  The three relevant angles are
$\varphi_{cb,c}    = \lambda_c-2\lambda_b+\varpi_c$,
$\varphi_{cb,b}    = \lambda_c-2\lambda_b+\varpi_b$, and
$\varphi_{cb} = \varphi_{cb,b}-\varphi_{cb,c} = \varpi_b-\varpi_c$.
For each of these angles, we measure the amplitude as half the difference
between the maximum and minimum values obtained during each $\sim$821-year
simulation.  As expected, we find that for both two- and three-planet fits to
the real data set and for every bootstrapped data set, all three angles librate
about 0$^{\circ}$.  We use the subscript ``cb'' to indicate the angles relating
the longitudes of planets ``c'' and ``b.''  In \S~\ref{4plcf}, we will use the
subscripts ``be'' and ``ce'' to indicate the corresponding angles relating the
longitudes of planets ``b'' and ``e,'' and ``c'' and ``e,'' respectively.

\begin{figure}
\includegraphics[angle=-90,scale=0.60]{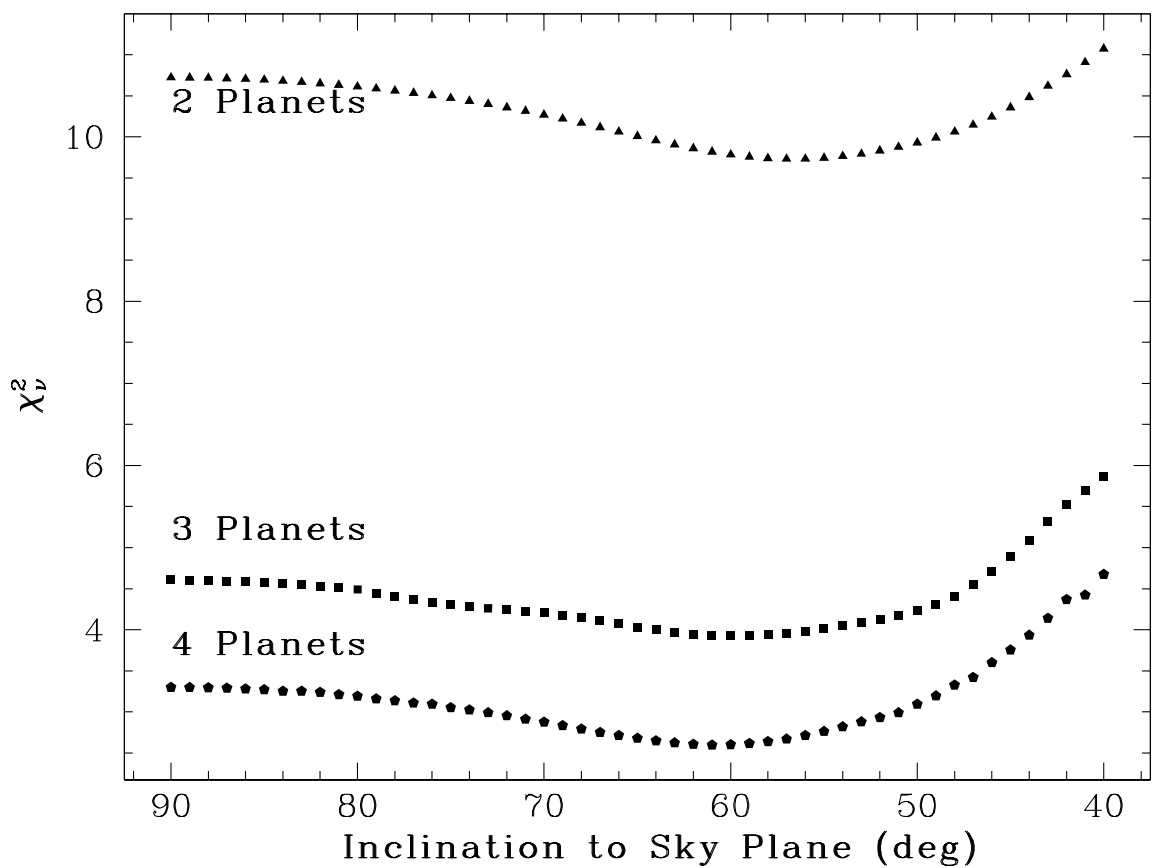}
\caption{For GJ~876, $\chi_{\nu}^2$ as a function of the coplanar system's
inclination to the plane of the sky.  The different types of points indicate
how many planets are assumed to be in the system.  Triangles are for two
planets, squares are for three planets, and pentagons are for four planets.
The eccentricity of ``d'' is allowed to float for all of the three- and
four-planet fits used to make this figure.  Note that the minimum is better
constrained with the addition of each planet.}
\label{chivsinc}
\end{figure}

As in R05, we inclined the coplanar system to the line-of-sight.  We set the
inclination relative to the plane of the sky of all the companions to values
from 90$^{\circ}$ to 40$^{\circ}$ in decrements of 1$^{\circ}$.  We hold all the
longitudes of the ascending node fixed at 0$^{\circ}$.  We fit for all the
remaining parameters.  Finally, we examine $\chi_{\nu}^2$ as a function of the
system's inclination.  Figure~\ref{chivsinc} shows the results. 

The minimum in $\chi_{\nu}^2$ as a function of the system's inclination is in
the range 57\,--\,61$^{\circ}$.  We find that the location of the minimum has
a small dependence on the number of companions in the system.
For two planets, we find $i=57\pm5.8^{\circ}$.
For three planets, we find $i=59\pm3.2^{\circ}$.
For four planets, we find $i=61\pm2.4^{\circ}$.
(The results for four-planet fits are presented in \S~\ref{4plcf}).  The error
estimates on the coplanar inclination are derived from 100 bootstrap trials of
the RV data set.  For each bootstrap RV set, we use the best-fit parameters to
the real RVs at each inclination value as the initial guess.  For each trial, we
find the minimum in $\chi_{\nu}^2$ as a function of the system's inclination.
The uncertainties in the fitted inclination of the system are the standard
deviations of the locations of the minimum in $\chi_{\nu}^2$.  Our best three-
and four-planet dynamical fits indicate a system with an invariable plane with
an inclination relative to the plane of the sky of $\sim59^{\circ}$, and we
adopt this inclination as a starting point for studying the fitness of models
with mutually inclined orbits.

Starting from the coplanar, $i=59^{\circ}$, two-planet fit, we examine the
effect of a mutual inclination between companions ``b'' and ``c.''  We do this
by fitting for the inclinations of ``b'' and ``c'' ($i_b$ and $i_c$) and the
longitude of the ascending node of ``c'' ($\Omega_c$).  Since the system is
rotationally invariant about the line-of-sight, we leave the longitude of the
ascending node of ``b'' ($\Omega_b$) fixed at 0$^{\circ}$.
We find $\chi_{\nu}^2=9.76$ (rms$=$5.704 m\,s$^{-1}$) for the coplanar fit, and
$\chi_{\nu}^2=9.91$ (rms$=$5.706 m\,s$^{-1}$) for the mutually inclined fit.
This ``improvement'' is not significant.  As we show below, fitting for the
mutual inclination of ``b'' and ``c'' in the case of four-planet configurations
results in a similar ``improvement'' in $\chi_{\nu}^2$.  For the two-planet
mutually inclined model, the fitted masses and inclinations are
$m_c=0.7211\,M_{\rm Jup}$, $m_b=2.3689\,M_{\rm Jup}$, $i_c=57.94^{\circ}$, and
$i_b=55.29^{\circ}$, and $\Omega_c=0.82^{\circ}$.  With the approximation that
$\Omega_c\sim0^{\circ}$, the invariable plane of this two-planet system has an
inclination relative to the plane of the sky of $56^{\circ}$.  The mutual
inclination for this system is 2.7$^{\circ}$.

For the three-planet model with $i=59^{\circ}$,
our best coplanar fit has $\chi_{\nu}^2=3.93$ (rms$=$3.625 m\,s$^{-1}$), and the
mutually inclined fit has $\chi_{\nu}^2=3.99$ (rms$=$3.620 m\,s$^{-1}$).  For the
mutually inclined fit, the invariable plane of the system has an inclination
relative to the plane of the sky of $59.2^{\circ}$.  For the mutually inclined
fit, we force planet ``d'' to be in the invariable plane determined by ``b'' and
``c.''  The mutual inclination between ``b'' and ``c'' in this configuration is
0.7$^{\circ}$.  Since a mutual inclination does not significantly improve the
three-planet fit, in Table~\ref{3pl_i=59} we show the best-fit parameters for
our best coplanar three-planet fit.

At $i=59^{\circ}$, the angles $\varphi_{cb,c}$, $\varphi_{cb,b}$, and
$\varphi_{cb}$ are all in libration for both the model fit (in
Table~\ref{3pl_i=59}) and for all of the bootstrap trials.  For the fit in
Table~\ref{3pl_i=59}, these amplitudes are
$\varphi_{cb,c} = 4.5\pm0.7^{\circ}$,
$\varphi_{cb,b} = 13.1\pm3.2^{\circ}$, and
$\varphi_{cb} = 11.5\pm3.7^{\circ}$.
Note that the libration amplitudes for the three-planet fits presented in this
section are smaller than for the three-planet fits presented in R05.

The critical angles show a complicated evolution for the coplanar three-planet
fit with $i=59^{\circ}$.  The first angle, $\varphi_{cb,c}$, has peaks at (listed
with decreasing power) 1.58, 45.67, and 8.69 years.  The second angle,
$\varphi_{cb,b}$, has peaks at 8.69, 1.58, and 45.74 years.  Finally, the third
angle, $\varphi_{cb}$, has two peaks at 8.69 and 45.67 years.  These
periodicities are also present in the evolution of the eccentricities of the
two outer planets.  The eccentricities of all three planets show very regular
secular evolution.

\begin{deluxetable}{lccc}
\tabletypesize{\scriptsize}
\tablecaption{Three-Planet coplanar fit for GJ~876 with $i=59^{\circ}$}
\tablewidth{0pt}
\tablehead{Parameter & Planet d & Planet c & Planet b\\}
\startdata
$P$ (days)                           & $1.937781\pm0.000024$     & $30.0880\pm0.0091$    & $61.1139\pm0.0084$\\
$m$\tablenotemark{a}                 & $6.78\pm0.51\,M_{\oplus}$ & $0.7175\pm0.0042\,M_{\rm Jup}$ & $2.2743\pm0.0059\,M_{\rm Jup}$\\
$a$\tablenotemark{a} (AU)            & $0.02080665\pm0.00000018$ & $0.129590\pm0.000026$ & $0.208301\pm0.000019$\\
$K$ (m\,s$^{-1}$)                    & $6.60\pm0.47$             & $88.72\pm0.52$        & $213.86\pm0.55$\\
$e$                                  & $0.257\pm0.070$           & $0.25493\pm0.00080$   & $0.0292\pm0.0015$\\
$\omega$ ($^{\circ}$)                & $229\pm28$                & $48.67\pm0.82$        & $50.7\pm4.5$\\
MA ($^{\circ}$)                      & $358\pm29$                & $295.0\pm1.0$         & $325.4\pm4.6$\\
offset (m\,s$^{-1}$)                 & \multicolumn{3}{c}{$50.7\pm0.4$}\\
Fit Epoch (JD)                       & \multicolumn{3}{c}{2450602.093}\\
$\chi_{\nu}^2$                        & \multicolumn{3}{c}{3.9280}\\
RMS (m\,s$^{-1}$)                    & \multicolumn{3}{c}{3.6254}\\
$\varphi_{cb,c}$                     & \multicolumn{3}{c}{$4.5\pm0.7^{\circ}$}\\
$\varphi_{cb,b}$                     & \multicolumn{3}{c}{$13.1\pm3.2^{\circ}$}\\
$\varphi_{cb}$                       & \multicolumn{3}{c}{$11.5\pm3.7^{\circ}$}\\
\enddata
\tablenotetext{a}{Quoted uncertainties in planetary masses and semi-major
axes {\it do not} incorporate the uncertainty in the mass of the star}
\label{3pl_i=59}
\end{deluxetable}

\section{Residuals of the Three-Planet Fit}
\label{Res-3pl}

\begin{figure}
\includegraphics[angle=-90,scale=0.60]{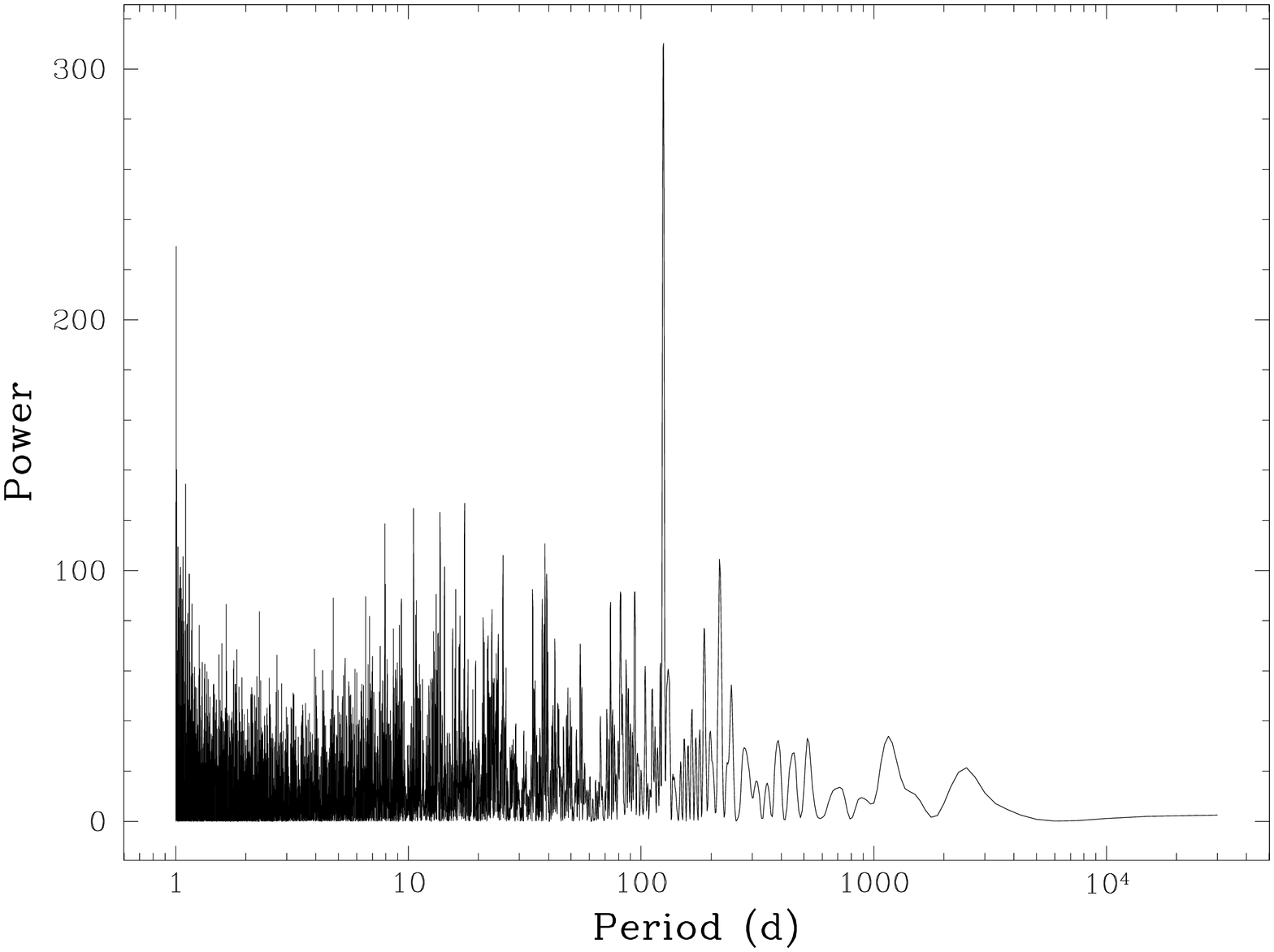}
\caption{Error-weighted periodogram of the residuals of the three-planet
coplanar fit, with $i=59^{\circ}$, to the RV data of GJ~876.}
\label{period59}
\end{figure}

\begin{figure}
\includegraphics[angle=-90,scale=0.60]{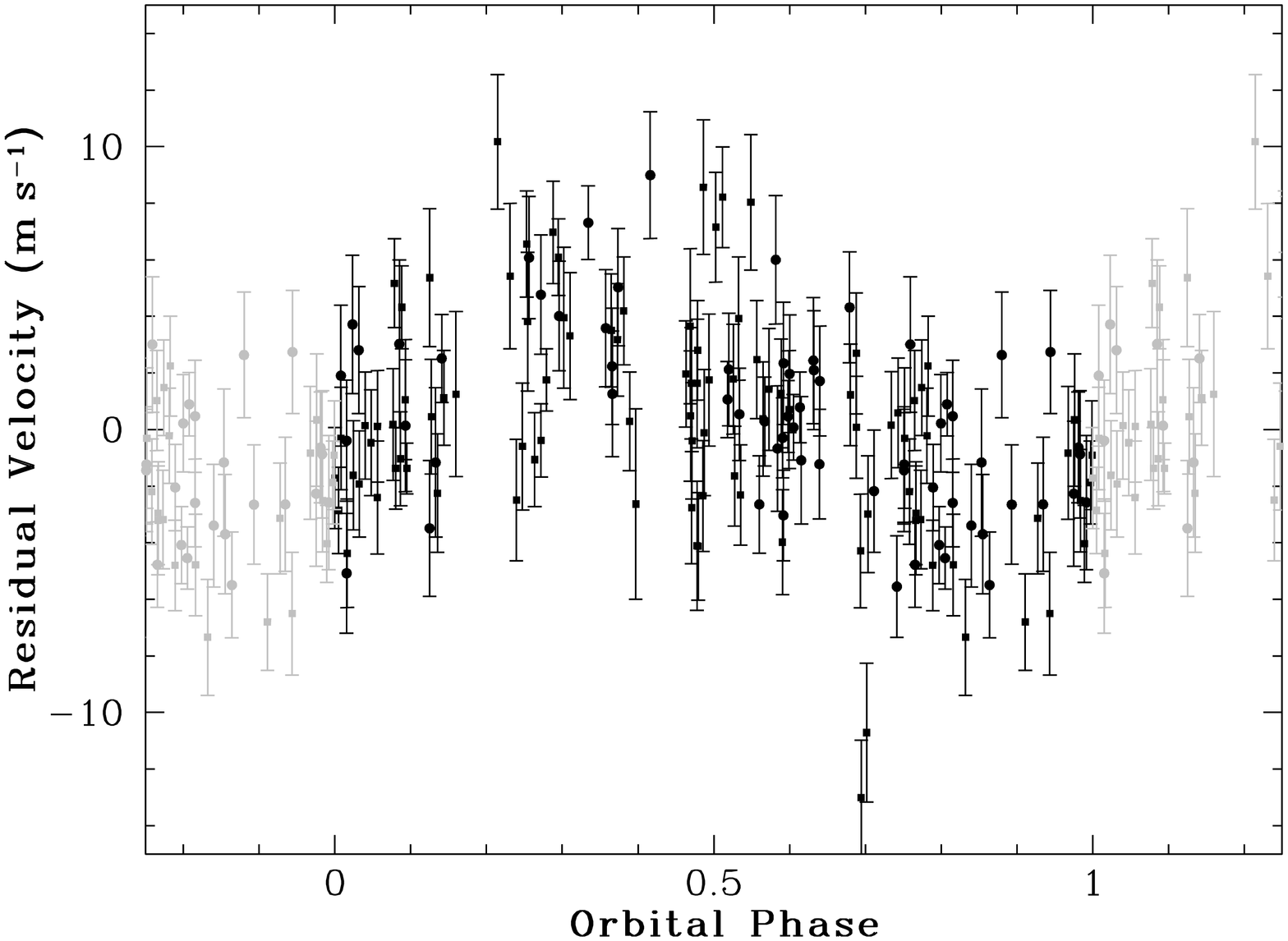}
\caption{Residuals of the three-planet coplanar fit, with $i=59^{\circ}$, to the
RV data of GJ~876.  The residuals have been folded at the period of the tallest
peak in the periodogram of the residuals.  The first 98 observations are shown
as squares.}
\label{fold59}
\end{figure}

We performed an error-weighted Lomb-Scargle periodogram \citep{GB87} analysis on
the residuals of the coplanar three-planet fits for both $i=90^{\circ}$ and
$i=59^{\circ}$.  Both periodograms show a significant peak at $\sim$125 days.
Figure~\ref{period59} shows the periodogram for the case $i=59^{\circ}$.  The
periodogram for the case $i=90^{\circ}$ looks very similar.  For $i=90^{\circ}$,
the peak raw power is 311.4.  For $i=59^{\circ}$, the peak raw power is 310.2.
Even if we allow for a mutual inclination between ``b'' and ``c'' and an overall
trend as in \citet{HARPS} (they actually subtract out a trend, due to the
geometrical effect of the star's proper motion, prior to fitting
the RVs), we still see a strong signal near 125 days in the periodogram of the
residuals.  For both values of $i$, folding the three-planet residuals at the
peak period shows a strong coherent sinusoidal signature.  Figure~\ref{fold59}
shows the folded residuals for the case $i=59^{\circ}$.  Examination of all the
inclined three-planet fits shows that the peak period in the residuals is at
125 days for all values of $i\geq40^{\circ}$.  For the 1000 fits used in
determining the uncertainties in Table~\ref{3pl_i=59}, the residuals of 941 of
them have a peak period near 125 days, with 12 harboring a marginally higher
alias peak with a period near 1.0 days.  The 125-day periodicity has been
adequately sampled at all phases only relatively recently.

We also examined subsets of the RV data where we took just the first $N$
observations, assumed $i=59^{\circ}$, and fit for the parameters of the three
known planets.  We took all values of $N$ from 98 to 162.  We then examined the
error-weighted Lomb-Scargle periodogram of the residuals for all values of $N$.
All the periodograms have a peak period near 125 days.  In Figure~\ref{fold59}
we show the folded residuals for the first 98 observations as squares.
Comparison between the folded residuals for the first 98 points and that for the
full RV set again suggests that the 125-day periodicity has been sampled at all
phases only recently.  The observed power at 125 days has grown monotonically
during the last decade, but the growth rate has been non-uniform.  The power
grows more quickly whenever we observe around the time when the fourth planet is
near the phases in its orbit when its orbital velocity has a large component
along the line-of-sight.  The power grows less quickly whenever we observe the
system when the orbital motion of the fourth planet is mostly perpendicular to
the line-of-sight.  We are constrained to observe near the time of the full
Moon, which impedes the efficiency of obtaining uniform phase coverage for
periodicities near 120 days.  As a consequence, there were several stretches
where we consistently observed near the phases when the fourth planet's orbital
motion was primarily perpendicular to the line-of-sight.  This circumstance is
one of the reasons why so many observations over 12 years were required prior to
reliably detecting the fourth periodicity in this system.

As an additional check on the reality of the 125-day signal, we ran a Monte
Carlo analysis.  We take the coplanar three-planet model at $i=59^{\circ}$ and
integrate forward to generate a synthetic RV curve.  We then generate a set of
1000 synthetic data sets by repeatedly sampling the synthetic RV curve at the
times of the Keck observations and adding Gaussian noise perturbations of
amplitude equal to the rms of the fit in Table~\ref{3pl_i=59}.  Then, for each
RV set, we fit for the three planets and examine the periodogram of the
residuals.  In not one instance did the power near 125 days exceed the value
observed for the real data.  Additionally, the peak period is ``near'' 125 days
in only one of the 1000 cases.  Thus, the false alarm probability (FAP) is
$<0.001$.

An alternative to the fourth planet hypothesis is that the 125-day periodicity
is due to rotation of the star itself.  The rotation period of GJ~876 is at
least $\sim40$ days, based on its narrow spectral lines and its low
chromospheric emission at Ca II H\&K \citep{detal98}.  R05 presented
photometric evidence of a rotation period of $\sim$97 days.  Our measurements
of the Mt.\ Wilson S activity index also confirms this rotation period.  Thus,
rotational modulation of surface features cannot explain the 125-day period in
the velocities.  The periodogram of our measured Mt. Wilson S activity index
after subtracting out a 14-year spot cycle has a peak period near 88 days.

\section{Four-Planet Fits}
\label{4plcf}

The 125-day periodicity in the residuals to our best coplanar three-planet fit
motivates exploration of the possibility that a fourth companion exists in the
system.  For the past several years, as we accumulated new RV measurements, we
monitored the quality of dynamical fits containing a fourth planet with
$P\sim125$ days.  From late 2005 through 2008, these fits preferred a fourth
planet with $e_e>0.2$, and produced system configurations which, when
integrated, were dynamically unstable.  Only recently, have the four-planet fits
resulted in systems which are stable on time scales exceeding a few hundred
million years (at least for coplanar configurations).  A fourth planet, with a
mass similar to Uranus, and on a stable, low-eccentricity orbit, has now
definitively emerged from the data.

A primary reason why the latest fits result in a stable system is that we have
only recently gathered enough observations to sample all phases of the 125-day
periodicity.  Figure~\ref{fold59} shows that most of the (small) phase gaps
have now been filled in.  Additionally, near the points with larger residuals,
we also have additional observations with smaller residuals.  As a result, the
LM routine no longer prefers a fourth planet with period $\sim$125 days with a
relatively large eccentricity.  Another reason why it has been difficult to
find stable four-planet fits with the fourth planet at 125 days is that the
region around the outer 2:1 MMR with companion ``b'' contains the boundary
between stable and unstable orbits for the two-planet system consisting of
only ``c'' and ``b'' \citep{RH07}.

\begin{figure}
\includegraphics[angle=-90,scale=0.60]{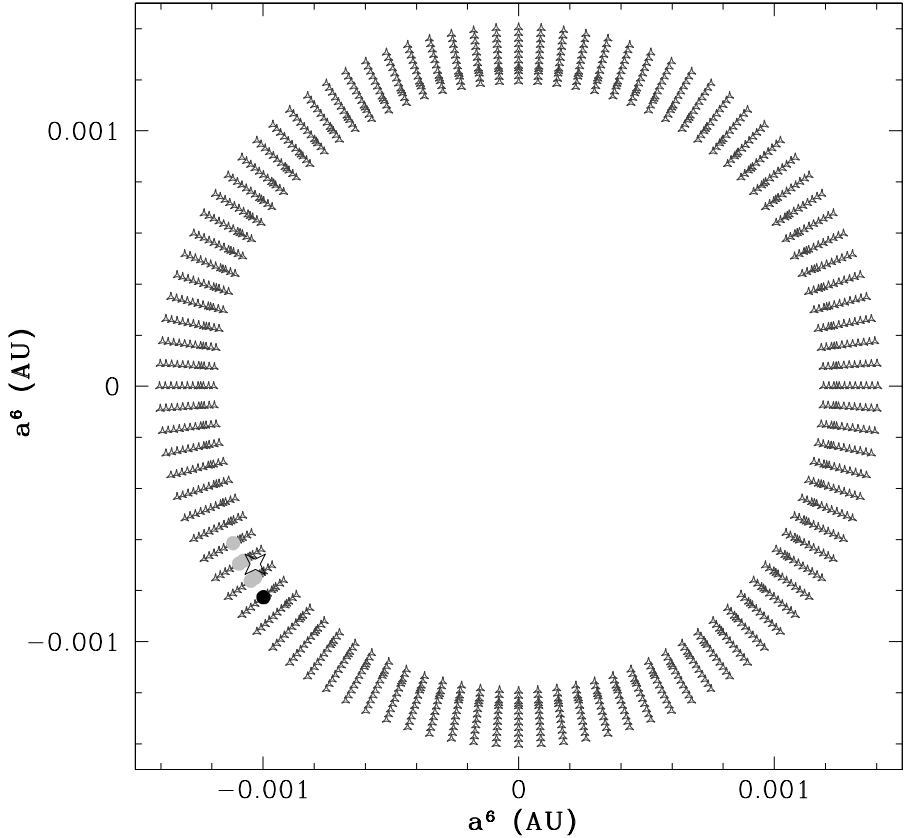}
\caption{Stability of massless test particles placed at equally spaced initial
longitudes on circular orbits with periods of 120\,--\,125 days in the plane of
the system based on the fit from Table~\ref{3pl_i=59}.  For clarity, we show the
astrocentric semi-major axes raised to the sixth power.  Unstable locations are
indicated with dark grey open symbols.  The single ``stable'' location is
indicated with a black filled-in symbol.  The fitted initial location of the
fourth planet from the fit in Table~\ref{4pl_i=59} is marked as an open star.
Stable locations under the assumption $e_d=0$ are indicated with light grey
filled-in symbols.
}
\label{stargate}
\end{figure}

To highlight the generally tenuous stability of planets with periods near 125
days, we performed simulations based on the fit in Table~\ref{3pl_i=59} in which
we added 1200 (massless) test particles on circular orbits in the plane of the
system with periods in the range 120\,--\,125 days every 0.5 days plus at the
fitted period for ``e.''  At each period we placed 100 test particles equally
spaced in initial longitude.  In agreement with \citet{RH07}, nearly all are
unstable on time scales $<10^6$ years.  Only one particle survived more than
$10^6$ years; however, it was also lost after 7.4 Myr.  It started with a period
of 122.5 days (astrocentric) at a mean anomaly of 219.6$^{\circ}$.  For this
long-term survivor the critical angle,
$\varphi_{\rm Laplace}=\lambda_c-3\lambda_b+2\lambda_e$, librates around
0$^{\circ}$ with an amplitude of $\Delta \varphi_{\rm Laplace}=22^{\circ}$ for the
entire time that the particle is stable.  The chaotic nature of the particle's
orbit becomes apparent when the libration amplitude of the Laplace angle
suddenly becomes unconstrained during the last few thousand years of its
evolution.  In Figure~\ref{stargate} we show the stability of the test particles
as a function of initial semi-major axis and longitude.  For clarity, we show
the astrocentric semi-major axes raised to the sixth power.  Unstable initial
locations are indicated with open dark grey symbols.  The single ``stable''
initial location, where a test particle survived 7.4 Myr, is indicated with a
filled-in black symbol.  With an open star, we also mark the approximate fitted
location for the fourth planet from the coplanar four-planet fit with
$i=59^{\circ}$ (for simplicity, we assume zero eccentricity for the fourth
planet --- the astrocentric eccentricity is actually 0.0448).  We also find that
if we assign a mass of a few Earth masses to these ``test particles,'' they tend
to survive for longer times.  The results strongly suggest that a resonant
mechanism is required for an object in this region to be stable.

Additionally, we find that the orbital parameters of the innermost planet,
``d,'' can influence the stability and chaotic nature of test particles placed
in the region near 125 days.  Assuming a circular orbit for ``d'' results in an
island of stability near the fitted location of planet ``e.''  Instead of
only a single particle which survives for 7.4 Myr, six particles survive for at
least 10 Myr in this region if $e_d$ is assumed to be zero.  As for the
long-lived particle for the case with $e_d>0$, all the ``stable'' particles are
in the Laplace resonance.
The six stable particles under the assumption $e_d=0$ are shown as filled-in
light grey symbols in Figure~\ref{stargate}.  We also used the Lyapunov
estimator in the SWIFT integration package \citep{SWIFT} to analyze the chaotic
nature of test particle orbits with periods near 125 days.  We find that the
most stable particles in this region have Lyaponov times in the range 10$^3$ to
10$^4$ years.

Following procedures similar to the ones used for the two- and three-planet
fits, we worked up four four-planet fits.
The best-fit coplanar inclination ($i=59^{\circ}$) with $e_d=0$ has
$\chi_{\nu}^2=2.8738$ (rms=3.0655 m\,s$^{-1}$).
Working from this fit, holding $e_d$ fixed, but fitting for a mutual inclination
between ``b'' and ``c'' gives $\chi_{\nu}^2=2.8322$ (rms=3.0219 m\,s$^{-1}$).
If instead we fit for $e_d$ and assume a  coplanar system, the best-fit
inclination is at $i=61^{\circ}$ and $\chi_{\nu}^2=2.5991$
(rms=2.9492 m\,s$^{-1}$).
Finally, fitting for $e_d$ and a mutual inclination between ``b'' and ``c''
results in $\chi_{\nu}^2=2.6098$ (rms=2.9303 m\,s$^{-1}$).  Since fitting for
$e_d$ and the mutual inclination between ``b'' and ``c'' results in a
significant improvement in $\chi_{\nu}^2$, we take this last model to represent
our current best fit to the present Keck RV data set.  However, simulating this
model as well as other mutually inclined models with similar values of
$\chi_{\nu}^2$ shows that the resulting systems are unstable on time scales of
$\lesssim1$ Myr.  Also, with $e_d$ allowed to float, the ``improvement'' in
$\chi_{\nu}^2$ between the coplanar and mutually inclined models is not
significant.  The F-Test probability that these two fits are statistically the
same is 0.88.  Thus, Table~\ref{4pl_i=59} shows the best-fit parameters for the
coplanar four-planet fit with $i=59^{\circ}$, our preferred inclination for the
system.  The resulting system is stable for at least 300 Myr, and other coplanar
fits with similar values of $\chi_{\nu}^2$ result in systems which are stable
for hundreds of millions of years up to at least 1 Gyr.

For completeness, we note that for our best-fit mutually inclined system, the
invariable plane of the system is inclined $59.5^{\circ}$ to the plane of the
sky, and the mutual inclination between ``b'' and ``c'' is 3.7$^{\circ}$.
Fitting for the inclinations of planets ``d'' and ``e'' does not improve
$\chi_{\nu}^2$ significantly.  This result for the inclination of ``d'' is in
agreement with the similar result by \citet{HARPS}.  The F-Test probability for
our best four-planet (mutually inclined) fit versus the three-planet coplanar
fit with $i=59^{\circ}$ and with a fitted $e_d$ is 0.0066.  This small
probability indicates that the four-planet model is significantly different from
the three-planet model.  It should be noted that for the coplanar case with
$i=59^{\circ}$ the fitted longitude of planet ``e'' is near the location of the
long-lived particle discussed above.

The inclination of planet ``e'' does have a significant effect on the stability
of the system based on our best mutually inclined fit.  It also influences the
libration amplitudes of the critical angles to be discussed below.  The orbit of
planet ``e'' must have an inclination and node such that the precession rates of
both the nodes and longitudes of periastron of the outer three planets result in
a relatively small libration amplitude for the critical angle of the Laplace
resonance, $\varphi_{\rm Laplace}=\lambda_c-3\lambda_b+2\lambda_e$.  In other
words, systems with a small forced inclination for ``e'' should be more stable
than systems with a large forced inclination -- through experimentation we find
this to be true.  The coplanar four-planet fits are generally more stable than
the mutually inclined versions since a small libration amplitude for
$\varphi_{\rm Laplace}$ depends only on the precession rates of the longitudes of
periastron of the outer three planets and not on the rates of precession of
their nodes.  For the mutually inclined four-planet fit, we performed a brief
search of the parameter space spanned by $i_e$ and $\Omega_e$ as well as $i_d$
and $\Omega_d$.  Experimentation shows that all four of these can influence the
stability (and evolution) of this chaotic system.
Also, the libration amplitudes of all the librating critical angles are smaller
than if we simply place ``e'' and ``d'' in the invariable plane determined by
``b'' and ``c.''  Since we only explored a small region of the parameter space,
and since we found several equally good fits to the RV data with various values
for the forced inclination of ``e,''  other, more stable systems with smaller
forced inclinations are likely to exist.  One emerging pattern from the search
of the parameter space is that smaller forced inclinations, smaller libration
amplitudes, and more stable systems occur more likely for models in which the
orbital plane of ``e'' is more closely aligned with that of ``b.''  This
pattern is also in rough agreement with the finding that coplanar systems are
generally more stable than mutually inclined ones.

As for the two- and three-planet fits, the critical angles involving ``c'' and
``b'' librate about $0^{\circ}$ for all our four-planet fits as well as for all
1000 bootstrap fits used in determining the uncertainties in
Table~\ref{4pl_i=59}.  However, their amplitudes are generally larger because of
the perturbing influence of the fourth planet.

\begin{figure}
\includegraphics[angle=-90,scale=0.60]{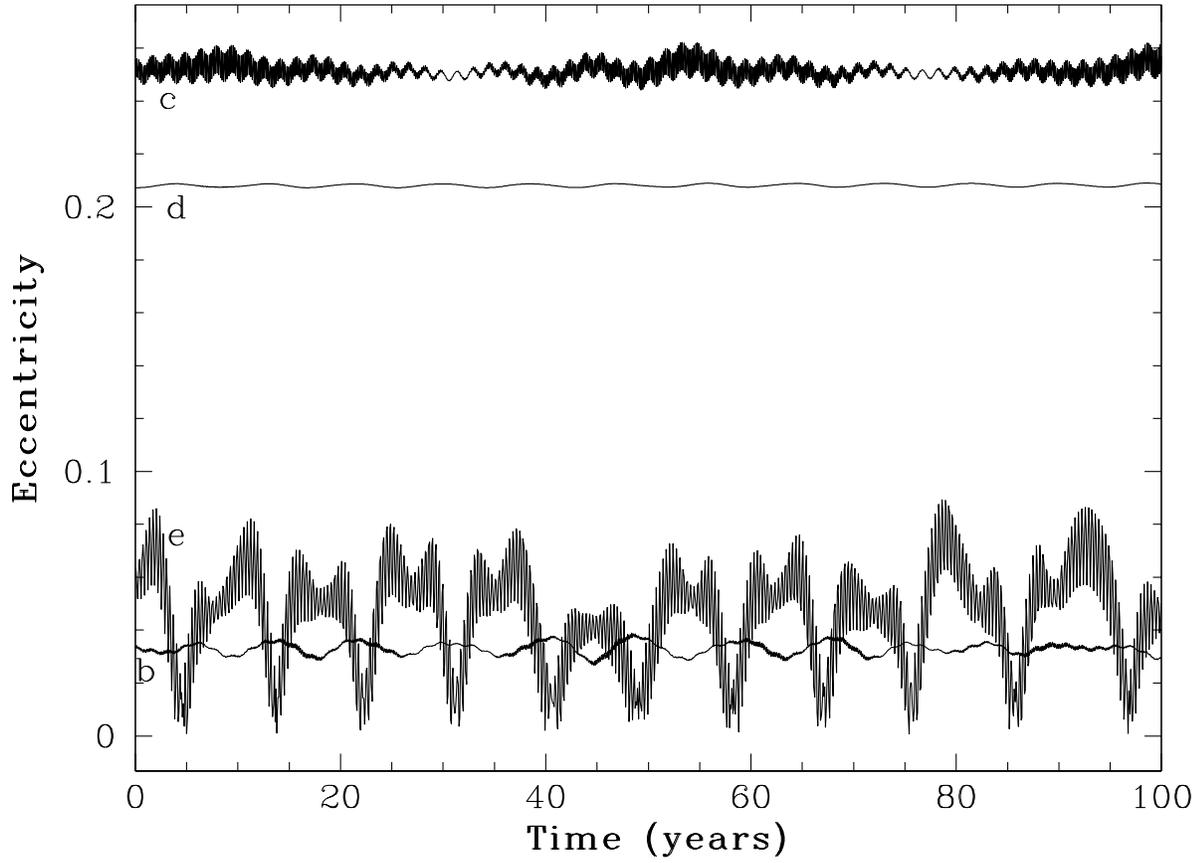}
\caption{Eccentricities of the four planets from the fit in
Table~\ref{4pl_i=59} over the first 100 years of an 821-year simulation.}
\label{4pl_i=59_1}
\end{figure}

\begin{figure}
\includegraphics[angle=-90,scale=0.60]{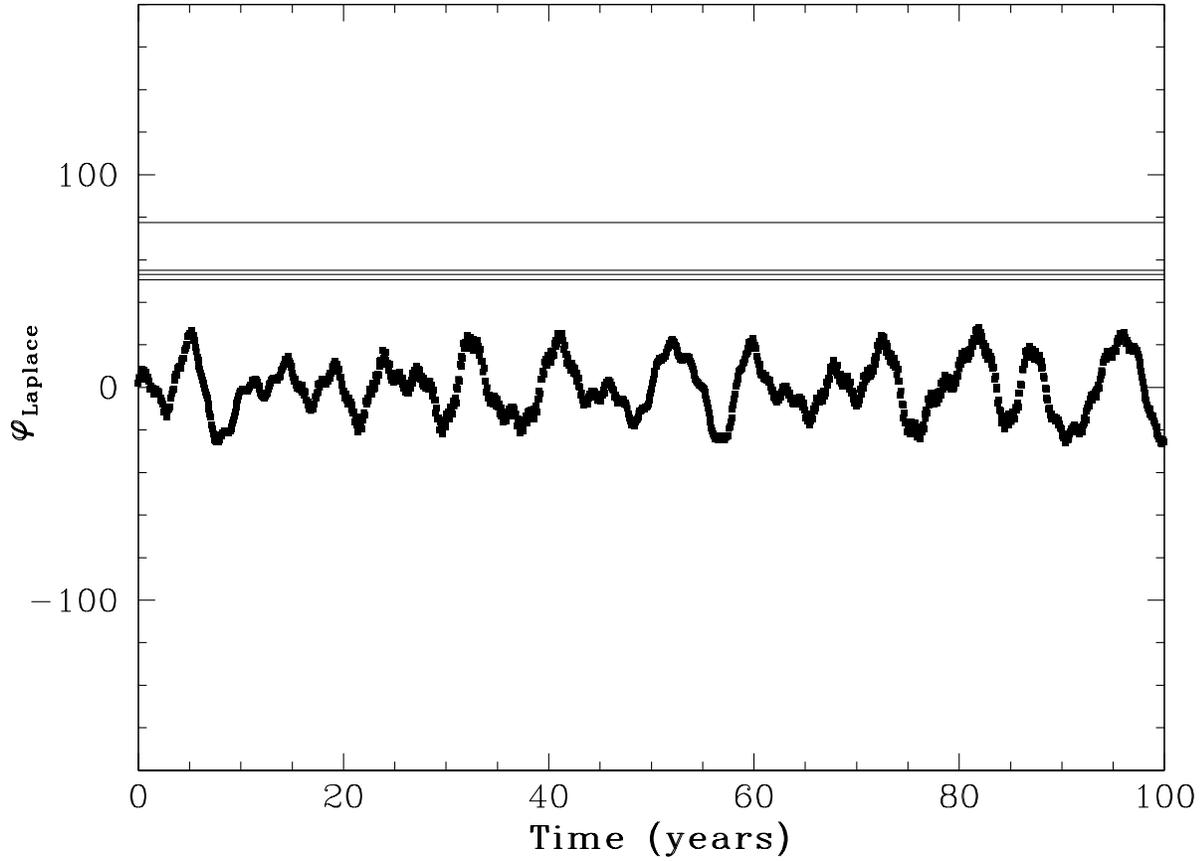}
\caption{Critical argument for the Laplace resonance for planets c, b, and e
over the first 100 years of an 821-year simulation.  The simulation is based on
the parameters from Table~\ref{4pl_i=59}.  Horizontal lines indicate the
observed amplitude for this simulation when it is extended to 10$^x$ years for
$x=$5, 6, 7, and 8, respectively.  The amplitude grows stochastically with
time.}
\label{4pl_i=59_5}
\end{figure}

Since the $\sim$125-day period is near the 2:1 MMR with planet ``b,'' the 4:1
MMR with planet ``c,'' and the Laplace resonance with both ``b'' and ``c,'' it
is informative to examine all the relevant critical angles associated with these
resonances as well as those associated with the resonances involving only
``b'' and ``c.''  For the 2:1 MMR and the linear secular coupling between ``b''
and ``e,'' the relevant angles are
$\varphi_{be,b}    = \lambda_b-2\lambda_e+\varpi_b$,
$\varphi_{be,e}    = \lambda_b-2\lambda_e+\varpi_e$, and
$\varphi_{be} = \varphi_{be,e}-\varphi_{be,b} = \varpi_e-\varpi_b$.
For the 4:1 MMR between ``c'' and ``e,'' there are 4 relevant angles:
$\varphi_{ce0}    = \lambda_c-4\lambda_e+3\varpi_c$,
$\varphi_{ce1}    = \lambda_c-4\lambda_e+2\varpi_c+\varpi_e$,
$\varphi_{ce2}    = \lambda_c-4\lambda_e+\varpi_c+2\varpi_e$, and
$\varphi_{ce3}    = \lambda_c-4\lambda_e+3\varpi_e$.
The subscripts above correspond to the multiplier in front of $\varpi_e$.
There is also, the critical angle involving just the longitudes of periastron
of ``c'' and ``e'':
$\varphi_{ce} = \varpi_e-\varpi_c$.
Finally, the critical angle for the Laplace resonance is
$\varphi_{\rm Laplace} = \lambda_c-3\lambda_b+2\lambda_e$.
Table~\ref{angles} lists the libration amplitudes of all these critical angles
for the fit in Table~\ref{4pl_i=59}.  A ``C'' in the amplitude column indicates
that the angle circulates (at least once) during the $\sim$821-year simulation. 
The time evolution of the planetary eccentricities and the Laplace angle for the
fit in Table~\ref{4pl_i=59} are shown in Figures~\ref{4pl_i=59_1} and
\ref{4pl_i=59_5}, respectively.  The secular evolution of the eccentricities of
planets ``b,'' ``c,'' and ``d'' is apparently slightly less regular as a result
of the addition of planet ``e,'' of which the eccentricity evolution is more
chaotic than that of the other three planets.  Additionally, the evolution of
all the librating angles in stable four-planet fits appears chaotic; however,
the resulting systems are stable for at least several hundreds of millions of
years.

The addition of the low-mass planet ``e'' to the system model has only a modest
effect on the dynamical characterization of the orbits of ``b'' and ``c.''
$\Delta\varphi_{cb,c}$ is increased from $4.5^{\circ}\pm0.7^{\circ}$ to
$5.74^{\circ}\pm0.85^{\circ}$, and $\Delta\varphi_{cb}$ increases from
$11.5^{\circ}\pm3.7^{\circ}$ to $22.5^{\circ}\pm4.8^{\circ}$.   (Additionally,
their evolution appears less regular with the addition of ``e.'')  In addition,
the 4:1 argument, $\varphi_{ce0}$, is observed to librate with
$\Delta\varphi_{ce0}=83^{\circ}\pm25^{\circ}$, and the 2:1 argument
$\varphi_{be,b}$ librates with $\Delta\varphi_{be,b}=36^{\circ}\pm13^{\circ}$.
Most significantly, the system also participates in the three-body resonance,
with $\Delta\varphi_{\rm Laplace}=40^{\circ}\pm13^{\circ}$.

Libration of $\varphi_{\rm Laplace}$ is required for stability.  Long-term
simulations based on bootstrap fits associated with four-planet coplanar fits
indicate that systems for which $\varphi_{\rm Laplace}\gtrsim140^{\circ}$ are
unstable on time scales $<2$ Myr.  Other systems with large libration
amplitudes for $\varphi_{\rm Laplace}$, $141.4^{\circ}$, $129.1^{\circ}$, and
$119.4^{\circ}$, are stable for only 25.9 Myr, 33.6 Myr, and 196.8 Myr,
respectively.  This suggests that the stability of the system also hinges on the
libration amplitude of the critical angle for the Laplace resonance.

\begin{figure}
\includegraphics[scale=0.35]{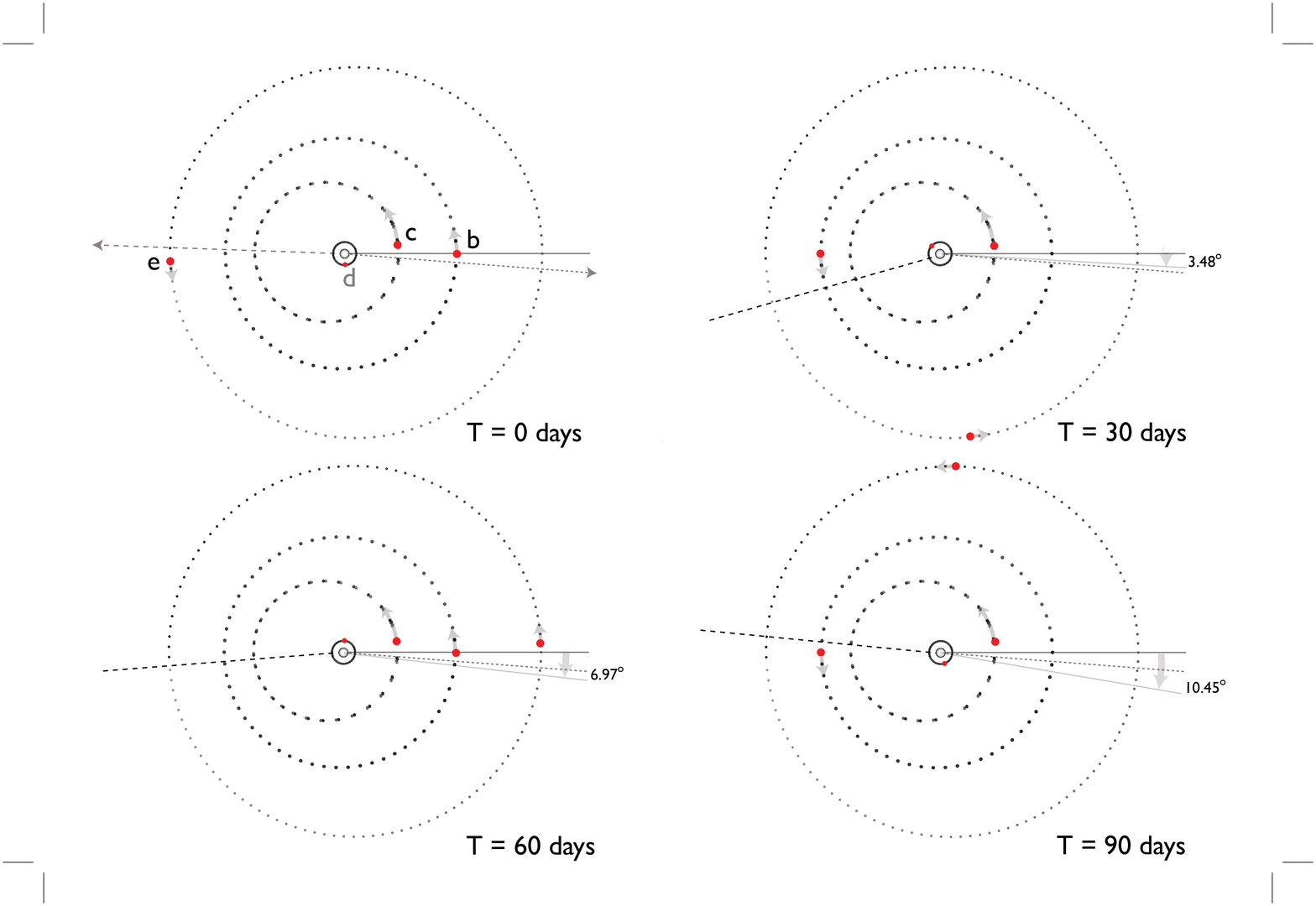}
\caption{Four configuration snapshots of the GJ~876 planetary system.  Each
panel shows the positions of the four planets at 120 successive one-day
intervals starting from JD=2450608.093 (T=0 days), with the orbital positions
at the listed times given by red dots.  The diagrams are drawn in a frame that
rotates to match the mean orbital precession of planet ``b,'' which amounts to
$-10.45^{\circ}$ over the 90 days shown.  Planet ``b'''s apsidal line coincides
with the x-axis.  The apses for planets ``c'' and ``e'' are drawn with
smaller-dashed and larger-dashed lines, respectively.}
\label{geometry}
\end{figure}

Figure~\ref{geometry} charts the planetary positions over 120 one-day intervals
starting from JD 2450608.093 in a frame rotating at planet ``b'''s precession
rate, $\langle \dot{\varpi_b} \rangle=-0.116165^{\circ}$\,d$^{-1}$.   Four
snapshots, corresponding to successive half-orbits for planet ``b'' are shown.
The plots trace the system's 1:2:4 commensurability at a time when the
osculating eccentricity of ``e,'' at $e_e\sim0.064$, is relatively high, and
when $\varpi_b-\varpi_e\sim180^{\circ}$.   During this phase, triple
conjunctions occur when planet ``e'' is near apastron.

For the dynamical configuration given in Table~\ref{4pl_i=59}, numerical
integration indicates that the periastron longitudes $\varpi_c$ and $\varpi_b$
regress more quickly than does $\varpi_e$.  Over a timescale of several years,
the three apses approach alignment, angular momentum is transferred to ``e,''
and the eccentricity of ``e'''s orbit declines nearly to zero.  During this
low-eccentricity phase, the arguments $\varphi_{be}$ and $\varphi_{ce}$ both
circulate through $2\pi$.  The complicated dynamical environment for ``e''
results in dramatic changes in its eccentricity over readily observable time
scales.  Our fit to the data indicates that during the time span of the RV
observations, the eccentricity of ``e'' has varied from $e_e\sim 0$ to
$e_e\sim0.075$ (see Figure~\ref{4pl_i=59_1}).

\begin{deluxetable}{lcccc}
\tabletypesize{\scriptsize}
\tablecaption{Four-Planet coplanar fit for GJ~876 with $i=59^{\circ}$}
\tablewidth{0pt}
\tablehead{Parameter & Planet d & Planet c & Planet b & Planet e\\}
\startdata
$P$ (days)                           & $1.937780\pm0.000020$     & $30.0881\pm0.0082$    & $61.1166\pm0.0086$      & $124.26\pm0.70$\\
$m$\tablenotemark{a}                 & $6.83\pm0.40\,M_{\oplus}$ & $0.7142\pm0.0039\,M_{\rm Jup}$ & $2.2756\pm0.0045\,M_{\rm Jup}$ & $14.6\pm1.7\,M_{\oplus}$\\
$a$\tablenotemark{a} (AU)            & $0.02080665\pm0.00000015$ & $0.129590\pm0.000024$ & $0.208317\pm0.000020$ & $0.3343\pm0.0013$\\
$K$ (m\,s$^{-1}$)                    & $6.56\pm0.37$             & $88.34\pm0.47$        & $214.00\pm0.42$       & $3.42\pm0.39$\\
$e$                                  & $0.207\pm0.055$           & $0.25591\pm0.00093$     & $0.0324\pm0.0013$     & $0.055\pm0.012$\\
$\omega$ ($^{\circ}$)                & $234\pm20$                & $48.76\pm0.70$          & $50.3\pm3.2$          & $239\pm22$\\
MA ($^{\circ}$)                      & $355\pm19$                & $294.59\pm0.94$         & $325.7\pm3.2$         & $335\pm24$\\
offset (m\,s$^{-1}$)                 & \multicolumn{4}{c}{$51.06\pm0.30$}\\
Fit Epoch (JD)                       & \multicolumn{4}{c}{2450602.093}\\
$\chi_{\nu}^2$                       & \multicolumn{4}{c}{2.6177}\\
RMS (m\,s$^{-1}$)                    & \multicolumn{4}{c}{2.9604}\\
\enddata
\tablenotetext{a}{Quoted uncertainties in planetary masses and semi-major
axes {\it do not} incorporate the uncertainty in the mass of the star}
\label{4pl_i=59}
\end{deluxetable}

\begin{deluxetable}{lc}
\tabletypesize{\scriptsize}
\tablecaption{Libration Amplitudes of Relevant Critical Angles}
\tablewidth{0pt}
\tablehead{Angle & Amplitude ($^{\circ}$)\\}
\startdata
$\varphi_{cb,c}$ & $5.74\pm0.85$\\
$\varphi_{cb,b}$ & $21.9\pm4.2$\\
$\varphi_{cb}$ & $22.5\pm4.8$\\
$\varphi_{be,b}$ & $36\pm13$\\
$\varphi_{be,e}$ & C\\
$\varphi_{be}$ & C\\
$\varphi_{ce0}$ & $83\pm25$\\
$\varphi_{ce1}$ & C\\
$\varphi_{ce2}$ & C\\
$\varphi_{ce3}$ & C\\
$\varphi_{ce}$ & C\\
$\varphi_{\rm Laplace}$ & $40\pm13$\\
\enddata
\label{angles}
\end{deluxetable}

\section{Are the Three Outer Planets Truly in a Laplace Configuration?}
\label{Laplace}

In the previous section we showed evidence for a fourth planet in the GJ~876
system, planet ``e,'' which participates in a 3-body resonance with planets
``c'' and ``b.''  Our preferred coplanar model for the GJ~876 planetary system
which is consistent with the RV data shows that the critical angle for the
Laplace resonance librates about 0$^{\circ}$ with amplitude $40^{\circ}$.  Given
the low RV half-amplitude for planet ``e,'' it is natural to question whether
there exists sufficient data to truly confirm the existence of the three-body
resonance between ``b,'' ``c,'' and ``e.''  In this section, we address the
questions concerning the reality of this dynamical configuration.  The analysis
in this section is another method to address the question of whether the
125-day period is actually due to a planet.

We performed a Monte Carlo simulation in which we took a three-planet model and
added a $\sim$125-day signal plus Gaussian noise.  The three-planet model is
just the three inner planets from the fit in Table~\ref{4pl_i=59}.
We added a Keplerian model with $P=120.713$ days, $m=14.629\,M_{\oplus}$,
$e=0.0448$, $\omega=251.36^{\circ}$, and MA=216.88$^{\circ}$ plus Gaussian noise
with rms=5.9696 m\,s$^{-1}$.  These (astrocentric) parameters are the result of
subtracting off the effect of the three inner planets from the fit in
Table~\ref{4pl_i=59} and fitting the residuals with a one-planet model (with
initial guess based on the parameters of the fourth planet in our preferred
four-planet fit) for which only the mean anomaly is allowed to float.  We
generated 1000 synthetic RV data sets and performed a four-planet Newtonian fit
to each data set.  We then integrated each fit for 300,000 days ($\sim$821
years), and examined the libration amplitude of the critical angle for the
Laplace resonance.  Note (again) that the synthetic RVs are based on a Newtonian
three-planet configuration plus a Keplerian signal plus Gaussian noise.  Thus,
we may expect that not too many fits should result in systems with a libration
amplitude comparable to the system based on our preferred four-planet model.
We find that only 6 simulations have a libration amplitude $<40^{\circ}$, and 29
have a libration amplitude $<53^{\circ}$, the one-sigma uncertainty on the
Laplace angle libration width.  Thus, most of these synthetic RVs result in
systems which are not comparable to the system based on our preferred
four-planet model.  These results suggest that we can constrain the libration
amplitude of the critical angle for the Laplace resonance.  The results also
indicate that planet ``e'' is strongly interacting with ``b'' and ``c.''  This
is in good accord with the eccentricity evolution shown in
Figure~\ref{4pl_i=59_1}.

Another Monte Carlo experiment we performed is to assume the system is deep in
the Laplace resonance and generate synthetic RVs to see how often we find
systems comparable to the system based on our preferred four-planet fit.  Among
the fits to the bootstrap RVs above, we found one which results in a system in
which the libration amplitude of the critical angle for the Laplace resonance is
$16.4^{\circ}$.  We take this as our model and add Gaussian noise with
rms=3.9604 m\,s$^{-1}$, the rms for the fit in Table~\ref{4pl_i=59}.  We
generate 1000 synthetic RV data sets and perform four-planet Newtonian fits to
them.  The 1000 fits are then integrated forward for $\sim821$ years.  We find
that 577 systems have a libration amplitude of the critical angle for the
Laplace resonance $<40^{\circ}$, 799 systems have this libration amplitude
$<53^{\circ}$, and none have the critical angle in circulation.  The largest
amplitude is $\sim130^{\circ}$.  Again, these results indicate that we are
constraining the libration amplitude of the critical angle of the Laplace
resonance.

Additionally, we examined the long-term stability of several bootstrapped and
synthetic systems for which the critical angle for the Laplace resonance is
not strictly librating.  We find that such systems are unstable on timescales
$<200$ Myr.  Thus, our preferred coplanar fit to the current data set shows a
system which is participating in a Laplace resonance, and this configuration
appears to be required for long-term stability.

As noted in Figure~\ref{4pl_i=59_5}, the libration amplitude for the critical
angle for the Laplace resonance grows stochastically for long-term
simulations performed up to at least 1 Gyr.  If this diffusion occurs in the
actual system and if we assume that the system were damped down initially, then
the currently observed libration amplitude for the critical angle for the
Laplace resonance may give a rough estimate of the age of the system.

\section{Are There More Bodies in the System?}
\label{morepls}

\citet{RH07} used massless test particles to show that, except for the
innermost region, the entire habitable zone is unstable for small planets
because of the perturbing influence of just planets ``c'' and ``b.''  Based on
their work, it is expected that interior to the orbit of ``e'' there may be a
small region of stability just exterior to the orbit of ``d.''  Exterior to the
orbit of ``e,'' we expect the region out to the location of the 2:1 MMR with
``e'' to be unstable.  Beyond this location, we expect small planets on
low-eccentricity orbits to be stable except possibly around the location(s) of
the 3:1 MMR (and perhaps the 4:1 MMR) with ``e.''

We have performed a similar analysis based on the parameters in
Table~\ref{4pl_i=59}.  In fact, due to the strong interactions among planets
``e,'' ``b,'' and ``c,'' we find that particles placed exterior to the orbit of
``d'' are unstable out to beyond the location of the outer 5:2 MMR with ``e.''
This result is similar to that found by \citet{RL00} in
which the interactions between the outer planets in the $\Upsilon$ Andromedae
planetary system clear out a region much larger than what would be cleared out
if only the outermost planet were present.

The analyses above were done with test particles placed on circular orbits
starting on the $x$-axis.  They suggest that the region among the planets is
unstable.  This is also suggested by the inherently chaotic nature of the
four-planet system.  A more thorough exploration of the initial positions of
test particles can be used to find among the planets small islands of stability
which are protected by some resonant mechanism.  \citet{HARPS} postulated that
a small planet could exist with a period near 15 days.  Such a planet is not
detectable with the current RV data set(s).  The addition of the fourth planet,
``e,'' may have a significant effect on the stability of such a planet.
However, we explored the possibility that such a planet could be participating
in a four-body resonance with the three outer planets.  Short-term simulations
of particles placed in this region do indeed show that a small fraction of them
are participating in the putative four-body resonance.  A few of these test
particles have the relevant critical angle,
$\phi=\lambda_f-4\lambda_c+5\lambda_b-2\lambda_e$, executing small amplitude
libration, $\Delta\phi\sim30^{\circ}$, about $\phi\sim\pm70^{\circ}$.  This
configuration would keep the potential fifth planet away from the line joining
the outer three planets and the star when triple conjunctions occur among
``c,'' ``b,'' and ``e.''  Again, the present RV data set(s) do not support
such a planet, but it would be interesting to discover a resonant configuration
which has no Solar System analogue.

\section{Summary and Discussion}
\label{dis}

Using 162 RV observations taken at the Keck I telescope with HIRES, we have
shown strong evidence for a fourth periodicity that is present in the radial
velocity of GJ~876.  This likely corresponds to a planet ``e'' with period
$\sim$124 days and a minimum mass of $\sim$13.4 $M_{\oplus}$.  This planet joins
the previously known planets ``d,'' ``c,'' and ``b'' with periods of 1.9378,
30.1, and 61.1 days, respectively.  Assuming that the four-planet system is
coplanar, we have shown that the inclination of the system relative to the plane
of the sky is $\sim59^{\circ}$.  Our best self-consistent fit shows that the
period and mass of the new companion are 124.2 days and 15.4 $M_{\oplus}$.
The value of $\chi_{\nu}^2$ above 1.0 for our fit in Table~\ref{4pl_i=59}
suggests that the stellar jitter may be at the 1\,--\,2 m\,s$^{-1}$ level and
holds out the possibility of additional low-amplitude objects in the system
which may be revealed with additional observations.

We have also shown that this new planet likely participates in a complicated
set of resonances involving planets ``c'' and/or ``b.''  Simulations indicate
that it is in a 2:1 MMR with planet ``b,'' a 4:1 MMR with planet ``c,'' and
a Laplace resonance with both ``c'' and ``b.''  Comparison with prior
four-planet fits using fewer RV observations indicates that as more RVs have
been accumulated, the ``fitted'' libration amplitudes of the critical angles
associated with these resonances have generally decreased.  Additionally,
long-term simulations based on our fits indicate that the complicated dynamical
structure is necessary for the system's long-term survival.  It is important to
note that our current best-fit coplanar model may change as we and other groups
gather more RV data on this fascinating system.  Thus, the system may be deeper
in the resonances mentioned above.  Additionally, future RVs may eventually
constrain the mutual inclination between planets ``b'' and ``c.''

Phenomenal signal-to-noise and a decade-plus observational baseline combine to
make the GJ~876 system a touchstone for studies of planetary formation and
evolution.  The presence of well-characterized planet-planet resonances can
potentially allow a detailed connection to be made between the properties of
GJ~876's protoplanetary disk and the planetary system that eventually emerged.
Clearly, the system was prolific, both in terms of the mass and the total
number of planets produced.

A zeroth-order prediction of the core-accretion paradigm for giant planet
formation is that the frequency of readily detectable giant planets should
increase with both increasing stellar metallicity and with increasing stellar
mass (Laughlin et al.\ 2004, Ida \& Lin 2004, 2005). During the past decade,
both of these trends have been established observationally (see, e.g.
\citet{FV05} for a discussion of the metallicity trend and
\citet{Johnson08} and \citet{Bowler10} for discussions of the mass trend.)

Until recently, however, there appeared to be little evidence for the strong
expected planet-metallicity correlation among the handful of M-dwarf stars that
are known to harbor giant planets. In particular, the results of
\citet{Bonfils05} suggested that GJ~876 has subsolar metallicity. This
result naturally induces speculation that GJ~876's giant planets might be the
outcome of gravitational instability (e.g. Boss 2000) rather than core
accretion.

\citet{JA09} have recently provided an update to the metallicity calibration
developed by \citet{Bonfils05}.  The \citet{JA09} calibration indicates that
the planet-bearing M-dwarfs {\it do} appear to be systematically metal-rich,
suggesting that there is no breakdown of the planet-metallicity correlation
as one progresses into the red dwarf regime.  In particular, the new
calibration indicates $[{\rm Fe/H}]=0.39$; GJ~876 is more than twice as
metal-rich as the Sun.

A supersolar metallicity for GJ~876 dovetails with a history of
vigorously efficient planet formation, but there remain a number of
fascinating questions with regards to the formation and migration
mechanisms that permitted the final configuration to be assembled.
We are preparing an in-depth analysis of these issues, even as we
continue to gather more observations of this most eminently
productive star.

\acknowledgements

We sincerely thank Jacob Bean for refereeing and improving this work.
S.S.V. gratefully acknowledges support from NSF grants AST-0307493 and
AST-0908870, and from the NASA KECK PI program.  R.P.B. gratefully acknowledges
support from NASA OSS Grant NNX07AR40G, the NASA Keck PI program, the Carnegie
Institution of Washington, and the NAI, NASA Astrobiology Institute.  G.L.
acknowledges support from NSF grant AST-0449986.  N.H. gratefully acknowledges
support from NASA EXOB grant NNX09AN05G, the NASA Astrobiology Institute under
Cooperative Agreement NNA04CC08A at the Institute for Astronomy, University of
Hawaii, and a Theodore Dunham J. grant administered by Funds for Astrophysics
Research, Inc.  We also acknowledge the major contributions of fellow members of our previous
California-Carnegie Exoplanet team: Geoff Marcy, Debra Fischer, Jason Wright,
Katie Peek, and Andrew Howard, in helping us to obtain many of the pre-2008 RVs presented in this
paper. The work herein is based on observations obtained at the W. M.
Keck Observatory, which is operated jointly by the University of California and
the California Institute of Technology, and we thank the UC-Keck, UH-Keck, and
NASA-Keck Time Assignment Committees for their support.  We also wish to extend
our special thanks to those of Hawaiian ancestry on whose sacred mountain of
Mauna Kea we are privileged to be guests.  Without their generous hospitality,
the Keck observations presented herein would not have been possible.  This
research has made use of the SIMBAD database, operated at CDS, Strasbourg,
France.

\end{document}